\documentclass[aps,pra,twocolumn,nopacs,superscriptaddress]{revtex4}
\usepackage{graphicx}
\usepackage{amsmath}
\usepackage{amssymb}
\usepackage{amsfonts}
\usepackage{color}
\usepackage{bm}

\newcommand{\eqname}[1]{\label{eq:#1}}
\newcommand{\bgar}{\begin{eqnarray}}
\newcommand{\enar}[1]{\label{eq:#1}\end{eqnarray}}

\newcommand{\kk}{ {\bf k}}

\newcommand{\xx}{ {\bf x}}

\newcommand{\eq}[1]{(\ref{eq:#1})}

\newcommand{\Psihd}{\hat\Psi^\dagger}

\newcommand{\Psih}{\hat\Psi}

\newcommand{\ahd}{\hat a^\dagger}
\newcommand{\ah}{\hat a}
\newcommand{\bhd}{\hat b^\dagger}
\newcommand{\bh}{\hat b}
\newcommand{\Lambdahd}{\hat \Lambda^\dagger}
\newcommand{\Lambdah}{\hat \Lambda}

\begin{document}

\title{
Density correlations and dynamical Casimir emission of Bogoliubov phonons in modulated atomic Bose-Einstein condensates
}



\author{Iacopo Carusotto}
\affiliation{CNR-INFM BEC Center and Dipartimento di Fisica, Universit\`a di Trento, via Sommarive 14, I-38050 Povo, Trento, Italy}

\author{Roberto Balbinot}
\affiliation{Dipartimento di Fisica dell'Universit\`a di Bologna and INFN sezione di Bologna, Via Irnerio 46, 40126 Bologna, Italy}

\author{Alessandro Fabbri}
\affiliation{Departamento de Fisica Teorica and IFIC, Universidad de Valencia-CSIC, C. Dr.Moliner, 50, 46100 Burjassot, Spain}

\author{Alessio Recati}
\affiliation{CNR-INFM BEC Center and Dipartimento di Fisica, Universit\`a di Trento, via Sommarive 14, I-38050 Povo, Trento, Italy}
\affiliation{Physik-Department, Technische Universit\"at M\"unchen, D-85748 Garching, Germany}


\begin{abstract}
We present a theory of the density correlations that appear in an atomic Bose-Einstein condensate as a consequence of the dynamical Casimir emission of pairs of Bogoliubov phonons when the atom-atom scattering length is modulated in time.
Different regimes as a function of the temporal shape of the modulation are identified and a simple physical picture of the phenomenon is discussed. Analytical expressions for the density correlation function are provided for the most significant limiting cases. 
This theory is able to explain some unexpected features recently observed in numerical calculations of Hawking radiation from analog black holes. 
\end{abstract}


\maketitle

The dynamical Casimir effect~\cite{DCE} is a very general prediction of quantum field theory: whenever the boundary conditions and/or the dispersion law and/or the background of a quantum field are quickly varied in time, pairs of quanta are generated off the vacuum state by parametric amplification of zero-point noise.
The simplest and most celebrated example of dynamical Casimir effect was predicted for an optical cavity whose plane-parallel mirrors are made to rapidly oscillate in time along the cavity axis~\cite{Reynaud,lambrecht,Kardar}.
Despite the significant effort devoted to these fascinating effects, no experimental observation of the dynamical Casimir effect has yet appeared, the main reason being the difficulty in moving the mirror at a fast enough speed~\cite{lambrecht}.
Alternative schemes to modulate the effective optical length of a cavity on a very fast time-scale by acting on the refractive index of the cavity material have been proposed and shown to give a sizeable intensity of dynamical Casimir emission~\cite{law,PD,noi_bastard,noi_intersub,noi_DCE_atomi,DCE_atomi,circuit}. Experiments in this direction are in progress in several groups~\cite{delsing,Ciuti_nature}.

Since the original proposal by Unruh~\cite{unruh}, the advances in the field of the so-called analog models~\cite{analogy_book} have pointed out the possibility of simulating the physics of a quantum field on a generic curved space-time in table-top condensed-matter experiments: the propagation of elementary excitations in spatially and temporally inhomogeneous systems can in fact be recast in terms of a relativistic wave equation on an effective curved space-time.
In the simplest case of acoustic waves in a fluid, the space-time metric is fixed by the spatial and temporal profiles of the sound speed and of the flow velocity. 
Upon quantization of the resulting field theory, an analog dynamical Casimir effect is then expected to appear whenever the sound speed in a spatially homogeneous system is made to quickly vary in time, which in the language of the analogy corresponds to the expansion or contraction of the underlying universe~\cite{bec-analogy,gardiner}. On the other hand, in the presence of an acoustic horizon separating an upstream region of sub-sonic flow from a downstream one of super-sonic flow, the emission of analog Hawking radiation has been predicted~\cite{unruh,Helium,bec-analogy-hawk}.

As proposed in~\cite{Casimir_BEC_th_corr,noi_PRA_correlazioni}, a most promising way of experimentally investigating this physics involves the measurement of the correlation function of density fluctuations.
A recent {\em numerical experiment}~\cite{noi_NJP} has observed both the dynamical Casimir effect and the Hawking emission in a microscopic simulation of the dynamics of an atomic Bose Einstein condensate during and after the formation of an acoustic black hole.

After a series of papers~\cite{bec-analogy,gardiner,casimir_atomi_th,tozzo} focussed on the deposited energy and the spectrum of the emitted phonons, a number of theoretical works have started investigating the time-evolution of the correlation functions of atomic many-body systems after a sudden quench of some parameter, in particular the atom-atom interaction constant~\cite{Casimir_BEC_th_corr,calabrese,Cazalilla,Gritsev2,Demler}. 
In the meanwhile, several experiments have recently addressed the response of Bose gases
to a time-modulation of the atom-atom scattering length: the interest of most of these experiments was however concentrated on the energy deposited in the system by the perturbation~\cite{CasimirBEC_exp} and only provided qualitative information on the created density perturbation~\cite{CasimirBEC_exp2}.

As density correlations are becoming a standard observable in the experimental study of ultracold atomic gases~\cite{BECbook,manybodyBEC,correlations} we expect that the measurement of the density correlation pattern created by suitable perturbation sequences will soon provide a new powerful tool to investigate the microscopic properties of strongly correlated atomic gases and in particular of their elementary excitations. A few pioneering studies in this direction for the simplest case of a sudden quench have recently appeared~\cite{Gritsev2,Demler}.
 
It is remarkable to note that a similar strategy is presently under way in the completely different field of cosmology, where one is trying to extract information on the primordial inflation phase of the universe from the spectrum and the correlation function of the temperature fluctuations of the cosmological microwave radiation~\cite{wmap}. 
In the language of the analogy, the fast expansion of the universe in the inflation phase corresponds in fact to a fast temporal variation of the sound speed, which is indeed the result of a fast variation of the atom-atom scattering length~\cite{bec-analogy,gardiner,Casimir_BEC_th_corr}.

The present paper presents a comprehensive theoretical investigation of the different features that appear in the density correlation funcion of a spatially homogeneous Bose Einstein condensate as a consequence of a time-dependent atomic scattering length.
Calculations based the Bogoliubov theory of dilute condensates allow for a quantitative understanding of the different regimes as a function of the temporal profile of the scattering length modulation and provide a clear physical picture of the phenomenon in terms of the dynamical Casimir emission of entangled pairs of phonons. This theory fully confirms the numerical observations that was put forward in the original paper~\cite{noi_NJP}.

The paper is organized as follows. In sec.\ref{sec:system} we present the physical system under investigation and we briefly review the Bogoliubov approximation. The general theory of the density correlations that result from the dynamical Casimir emission is presented in Sec.\ref{sec:theory}. The following sections discuss the phenomenology in the most remarkable cases of an adiabatic transition (Sec.\ref{sec:adiabatic}), of a sudden jump (Sec.\ref{sec:jump}) and of a slow ramp (Sec.\ref{sec:slow}) of the atom-atom scattering length.
Analytical formulas valid in the hydrodynamic limit are presented in Sec.\ref{sec:hydro}. A comparison with the numerical results of~\cite{noi_NJP} is presented in Sec.\ref{sec:compMC}. The case of a quasi-periodic modulation of the scattering length is analyzed in Sec.\ref{sec:periodic} and a possible application to the measurement of temperature discussed.
The possibility of reinforcing the density correlation signal by mapping phase fluctuations into density ones is quantitatively studied in Sec.\ref{sec:trick}.
Conclusions are finally drawn in Sec.\ref{sec:conclu}.

\section{The physical system and the Bogoliubov description}
\label{sec:system}

We consider a spatially homogeneous Bose-Einstein condensate of atoms of mass $m$. The gas is assumed to be initially at rest in its thermal equilibrium state at a temperature $T$ and to have a density $n$. 
Atom-atom interactions are modeled by a repulsive, local interaction potential of scattering length $a>0$. The value of the scattering length is assumed to be constant in space but to have a non-trivial temporal dependence $a(t)$. In the remote past and future $t=\mp\infty$, it tends to a constant values.  
Experimentally, the possibility of tuning of the atom-atom interactions on a wide range has been demonstrated using magnetic and optical Feshbach resonances~\cite{feshbach}, as well as by modulating the lateral confinement of reduced dimensionality samples~\cite{BECbook}.

At all times, the system is assumed to be well within the dilute regime $n_0 a^3\ll1$ where the time-evolution of the condensate is accurately described by the so-called Bogoliubov approximation~\cite{BECbook,castin}. 
The time-evolution of the classical condensate wavefunction $\phi_0$ is described by the Gross-Pitaevskii equation,
\begin{equation}
i\hbar\frac{\partial\phi_0}{\partial t}=-\frac{\hbar^2\nabla^2}{2m}\phi_0+V(\xx)\,\phi_0+\frac{4\pi\hbar^2 a(t) }{m}\,|\phi_0|^2\,\phi_0.
\end{equation}
In the spatially homogeneous case $V(\xx)=0$ that we are considering here, the condensate wave function remains at all times constant in space and only acquires a time-dependent global phase, $\phi_0(\xx,t)=\sqrt{n}\,\exp{i\theta(t)}$. 

Within the Bogoliubov approximation~\cite{castin}, the fluctuations around the purely condensed state are described by a quadratic Hamiltonian in the $\Lambdah_\kk$ operators describing the non-condensed $\kk\neq 0$ plane-wave modes
\begin{multline}
\mathcal{H}=E_0+ \\
+\frac{1}{2} \sum_\kk 
\begin{array}{c}
(\Lambdahd_\kk , \Lambdah_{-\kk}) \\ \\ 
\end{array}
\left(\begin{array}{cc}
E_\kk+\mu(t) & \mu(t) \\ -\mu(t) & -E_\kk -\mu(t)
\end{array}\right)
\left(\begin{array}{c}
\Lambdah_\kk \\ 
\Lambdahd_{-\kk}
\end{array}\right)
\eqname{HamBogo}
\end{multline}
The $\Lambdah_\kk$ operators have a simple expression in terms of the momentum-space atomic field operators, $\Lambdah_\kk=N^{-1/2}\,\bhd_{\kk=0}\,\bh_{\kk}$. Here, $N$ is the total number of particles in the gas. In the following of the paper, we shall also make use of the real-space operators $\Lambdah(\xx)$ that are defined as the Fourier transform of the $\Lambdah_\kk$ operators~\cite{castin}.
The (instantaneous) chemical potential is defined in terms of the nonlinear interaction constant $g(t)=4\pi \hbar^2 a(t)/m$ as $\mu(t)=g(t)\,n$.
The kinetic energy of the $\kk$-mode is indicated as $E_\kk=\hbar^2 k^2/2m$. 

Neglecting the zero-point energy, the Hamiltonian \eq{HamBogo} can be recast for each instantaneous value of $a(t)$ into the canonical form
\begin{equation}
\mathcal{H}(t)=\sum_\kk \hbar \omega_\kk(t)\,\ahd_\kk\,\ah_\kk.
\eqname{HamBogo2}
\end{equation}
where the (bosonic) Bogoliubov operators $\ah_\kk$ ($\ahd_\kk$) respectively destroy (create) an elementary Bogoliubov excitation of momentum $\hbar \kk$ and energy 
\begin{equation}
\hbar\omega_\kk(t)=\sqrt{E_\kk(E_\kk+2\mu(t))}.
\eqname{omega}
\end{equation}
In terms of the atomic field operators, the (instantaneous) Bogoliubov operators $\ah_\kk$ have the following expression:
\begin{equation}
\ah_\kk =u_\kk(t)\,\Lambdah_\kk - v_\kk(t) \,\Lambdahd_{-\kk}.
\eqname{u_v}
\end{equation}
in terms of the (instantaneous) Bogoliubov coefficients $u_\kk(t)$ and $v_\kk(t)$,
\begin{equation}
u_\kk(t)\pm v_\kk(t) = \left( \frac{E_\kk}{\hbar \omega_\kk(t)} \right)^{\pm 1/2}~:
\end{equation}
as a consequence of the time-dependence of the scattering length $a(t)$, the Bogoliubov operators have an explicit time-dependence even in the Schr\"odinger picture of \eq{u_v}.
At each time, the (instantaneous) ground state $|g(t)\rangle$ of the Bogoliubov theory is defined by
\begin{equation}
\ah_\kk\,|g\rangle=0 \hspace{0.3cm} \forall \kk.
\end{equation}
The time-dependence of $|g(t)\rangle$ is the key responsible for the dynamical Casimir emission: when the scattering length is modulated at a fast rate, the system is not able to adiabatically follow the istantaneous ground state. Non-adiabatic processes then result in the creation of correlated pairs of excitations in the system out of the vacuum state~\cite{noi_bastard}. 
This point of view will be discussed in full detail in Sec.\ref{sec:jump}.

An alternative, yet equivalent picture of the dynamical Ca\-si\-mir emission can be obtained in the limiting case of a weak modulation $a(t)=a_0+\delta a(t)$ with $|\delta a(t) | \ll a_0$.
In analogy to the discussion of~\cite{law}, the dynamical Casimir emission can in this case be recast in terms of the following Hamiltonian
\begin{multline}
\mathcal{H}=\mathcal{H}_0+\frac{2\pi\hbar^2 n}{m}\,\delta a(t)\,\sum_\kk\,(u_\kk+v_\kk)^2 \\
\times (\ahd_\kk + \ah_{-\kk})(\ah_\kk+\ahd_{-\kk}).
\eqname{Ham_pert}
\end{multline}
Here, $\mathcal{H}_0$ is the Hamiltonian of the gas for a constant value of the scattering length $a_0$ and the effect of the weak modulation $\delta a(t)$ is to simultaneously excite pairs of entangled Bogoliubov particles with opposite momenta $\pm \kk$. In the language of nonlinear optics, such a process goes under the name of parametric down-conversion
\footnote{For the sake of completeness, it is important to note that a similar perturbation Hamiltonian 
$\delta\mathcal{H}=-
\sum_\kk\,\frac{\hbar^2\,k^2}{2m_0^2}\,\delta m(t)\,u_\kk v_\kk \left[ \ahd_\kk\,\ahd_{-\kk}+ \ah_{\kk}\,\ah_{-\kk}\right]$.
describes the phonon emission process that results from a modulation of the atomic mass $m(t)=m_0+\delta m(t)$. Such a time-modulation of the effective atomic mass appears, e.g., when the atoms are subjected to the periodic potential of a time-dependent optical lattice~\cite{CasimirBEC_exp}.}.
The most remarkable case of a periodic modulation of $a(t)$ will be discussed in Sec.\ref{sec:periodic}.

\section{The theoretical framework}
\label{sec:theory}

\subsection{The time-evolution of Bogoliubov operators}
Thanks to the quadratic nature of the Bogoliubov Hamiltonian \eq{HamBogo2}, the time-evolution of the Bogoliubov operators in the Heisenberg picture can be written in a closed form:
\begin{eqnarray}
\frac{d\ah_\kk}{dt} &=&-i\omega_\kk\,\ah_\kk + \left(\dot{u}_\kk\,v_\kk-u_\kk\,\dot{v}_\kk \right)\,\ahd_{-\kk}. \eqname{ak} \\ 
\frac{d\ahd_{-\kk}}{dt} &=&i\omega_{-\kk}\,\ahd_{-\kk} + \left(\dot{u}_\kk\,v_\kk-u_\kk\,\dot{v}_\kk \right)\,\ah_{\kk}. \eqname{ak2}
\end{eqnarray}
At each time $t$, the $u_\kk(t)$ and $v_\kk(t)$ functions are to be evaluated using the instantaneous value of the scattering length $a(t)$ according to \eq{u_v}. Dots indicate the derivative over time $t$. 

The first terms on the RHS of (\ref{eq:ak}-\ref{eq:ak2}) describe the trivial evolution of the $\ah_\kk$ and $\ahd_{-\kk}$ operators at frequencies $\pm \omega_\kk$ under the instantaneous Hamiltonian \eq{HamBogo2}.
The other terms take into account the dependence \eq{u_v} of the $\ah_\kk$ and $\ahd_{-\kk}$ operators on the instantaneous scattering length $a(t)$ via the time-dependence of the $u_\kk$ and $v_\kk$ Bogoliubov coefficients, and are responsible for the mixing of the $\ah_\kk$ and $\ahd_{-\kk}$ operators. These mixing terms are proportional to the rate at which $a(t)$ is varied in time.
Once $a(t)$ has approached its late-time limiting value, one is left with the trivial oscillation of the $\ah_\kk(t)$ and $\ahd_{-\kk}(t)$ operators at frequencies $\pm \omega_\kk$.

The relation between the $\ah_\kk(t)$ and $\ahd_{-\kk}(t)$ operators at a generic time $t$ to their initial values $\ah_\kk(0)$ and $\ahd_{-\kk}(0)$ before the excitation sequence
can be summarized as a pair of linear equations.
\begin{eqnarray}
\ah_\kk(t)&=&\left[C_{\kk,+}(t)\,\ah_\kk(0)+C_{\kk,-}(t)\,\ahd_{-\kk}(0)\right]
\eqname{jumpgen1} \\
\ahd_{\kk}(t)&=&\left[C_{\kk,+}^*(t)\,\ahd_{\kk}(0)+C_{\kk,-}^*(t)\,\ah_{-\kk}(0)\right], \eqname{jumpgen2}
\end{eqnarray}
whose coefficients $C_{\kk,\pm}(t)$ have to be computed by solving the pair of differential equations (\ref{eq:ak}-\ref{eq:ak2}).

Thanks to the thermal equilibrium hypothesis, the Bogoliubov modes are assumed to be initially uncorrelated and thermally occupied,
\begin{eqnarray}
\langle \ah_\kk(0) \ah_{-\kk}(0) \rangle&=&0 \eqname{aa} \\
\langle \ahd_\kk(0) \ah_{\kk}(0) \rangle&=&n_\kk^{{\rm th},0}=\frac{1}{\exp(\hbar \omega_\kk/k_B T)-1}. \eqname{a+a}
\end{eqnarray}

At late times $t>t_{\rm fin}$ after the end of the modulation, the Bogoliubov mode occupations 
\begin{multline}
n_\kk(t)=\langle \ahd_\kk(t)\,\ah_\kk(t) \rangle= \\ =\left( |C_{\kk,+}(t_{\rm fin})|^2+|C_{\kk,-}(t_{\rm fin})|^2 \right)\,n_\kk^{{\rm th},0}+\\
+|C_{\kk,-}(t_{\rm fin})|^2 \eqname{n_k_gen}
\end{multline}
remain constant in time, while the anomalous averages
\begin{multline}
\mathcal{A}_\kk(t)=\langle \ah_\kk(t)\,\ah_{-\kk}(t) \rangle = C_{\kk,+}(t_{\rm fin})\, C_{\kk,-}(t_{\rm fin}) \\ \times (2n_\kk^{{\rm th},0}+1)\,e^{-2i\omega_\kk\,(t-t_{\rm fin})} \eqname{A_k_gen}
\end{multline}
keep on oscillating at the frequency $2\omega_{\kk}$.

\subsection{The density correlation function}

In a homogeneous system the modulation of $a(t)$ has no effect on the average density that remains flat at all times,
\begin{equation}
n(\xx) = \langle \Psihd(\xx)\,\Psih(\xx) \rangle= n.
\end{equation}
On the other hand, the emission of entangled phonon pairs is clearly visible in the density correlation function
\begin{multline}
g^{(2)}(\xx,\xx')=\frac{1}{n^2}\,
\left\langle \Psihd(\xx)\,\Psihd(\xx')\,
\Psih(\xx')\,\Psih(\xx) \right\rangle=\\ 
=1+\frac{1}{nV}\sum_\kk \left[e^{i\kk(\xx'-\xx)} \left( (u_\kk+v_\kk)^2+1 \right) +\textrm{c.c.} \right]+ \\
+\frac{1}{nV}\sum_\kk (u_\kk+v_\kk)^2\,\left[e^{i\kk(\xx'-\xx)}\, \left( \langle \ah_\kk \ah_{-\kk} \rangle + \langle \ahd_\kk \ah_{\kk} \rangle \right) + \textrm{c.c.}\right]\\
=1+\frac{1}{nV}\sum_\kk\,e^{i\kk(\xx'-\xx)} \left[ (u_\kk+v_\kk)^2 \times \right. \\
\left. \times \left\langle (\ahd_\kk+ \ah_{-\kk})(\ah_\kk+ \ahd_{-\kk}) \right\rangle  -1\right]
 \eqname{G2anal}
\end{multline}
which involves a combination of Bogoliubov mode occupation \eq{n_k_gen} and anomalous average \eq{A_k_gen}. As we shall see in the following of the paper, the different time dependence of the two terms is responsible for qualitatively different features in the density correlation function.

\subsection{Effect of external trapping}

Before proceeding with the analysis of the density correlation function, it is important to briefly clarify the consequences of the spatial inhomogeneity of the gas in the presence of a trapping potential.

As the density profile of a trapped gas strongly depends on interactions, the modulation of the scattering length $a(t)$ may result into a macroscopic oscillation of the condensate shape~\cite{Cornell_collective} and even in its collapse when the scattering is tuned to a large and attractive value. This latter effect has been experimentally demonstrated in a remarkable way in the so-called Bose-nova experiments of~\cite{oscill_a}.

In the framework of the standard Bogoliubov theory~\cite{castin}, this physics is described by a term of the form
\begin{equation}
\delta\mathcal{H}_1=\frac{2\pi\hbar^2\,\delta a(t)}{m}\,\int\!d^3\xx\,|\phi_0(\xx)|^2\,\phi_0^*(\xx)\,\Lambdah(\xx)+\textrm{h.c.}.
\end{equation}
where $\phi_0(\xx)$ is the condensate wavefunction, normalized in a way that $\int\!d^3\xx\,|\phi_0(\xx)|^2=N$. 
As usual in quantum optics, an Hamiltonian term involving a single quantum field operator leads to a coherent excitation of the field, in our case a collective excitation of the condensate.
As expected, this term disappears in the case of a homogeneous condensate considered in the rest of the paper thanks to the spatial orthogonality of the $\Lambdah(\xx)$ operator to the condensate wavefunction $\phi_0(\xx)$,
\begin{equation}
\int\!d^3\xx\,\phi_0^*(\xx)\,\Lambda(\xx)=0.
\end{equation}
In the general case, the density fluctuation pattern can be isolated even in the presence of a strong collective excitation simply by subtracting out the deterministic component of the density modulation.

\section{Adiabatic limit}
\label{sec:adiabatic}

The mixing of the $\ah_\kk$ and $\ahd_{-\kk}$ operators is negligible $|C_{\kk,+}|\gg |C_{\kk,-}|\simeq 0$ in the so-called adiabatic limit where the time-evolution $a(t)$ takes place on a very slow time-scale as compared to $\omega_\kk$~\footnote{Note that we are here limiting ourselves a single-mode adiabaticity condition. For more general, global definitions of adiabaticity, one may refer to~\cite{Gritsev}.}.
As a consequence, the occupation $n_\kk$ of the Bogoliubov modes is constant in time and equal to its initial value $n_\kk^{{\rm th},0}$ and the anomalous averages $\mathcal{A}_\kk$ remains zero.

For a zero initial temperature $T=0$, the adiabatic condition is equivalent to stating that the evolution is slow enough for the system to remain in its ground state: at all times, the density correlation function exactly coincides with the static $T=0$ one for the instantaneous value of $a(t)$.
On the other hand, for a non-zero initial temperature $T>0$ even an adiabatic evolution is able to bring the system outside thermal equilibrium
\footnote{Rigorously speaking, this statement is valid only on time-scales that are short as compared to the characteristic time-scale for thermalization under the effect of the higher-order terms that are neglected in the Bogoliubov approximation \eq{HamBogo}. }:
while the population $n_\kk$ of each Bogoliubov mode is conserved during the adiabatic evolution, the instantaneous energy $\hbar\omega_\kk(t)$ of the mode has in fact a non-trivial time-dependence.
As a result, the population $n_\kk$ of the different Bogoliubov modes at late times is no longer described by a simple thermal condition of the form \eq{a+a}.

\section{Sudden jump}
\label{sec:jump}

\subsection{General formulas}

Simple expressions for the expectation values appearing in \eq{G2anal} can be obtained in the limit of a sudden variation of $a(t)$ from $a_1$ to $a_2$ on a time scale $\sigma_t$ much faster than the frequency $\omega_\kk$ of all the relevant modes, i.e. $\mu_{1,2} \sigma_t\ll 1$.
In this limit of a sudden {\em quench}~\cite{calabrese,Cazalilla,Gritsev2,kiss}, the evolution of the $\Lambdah_\kk(t)$, $\Lambdahd_{-\kk}(t)$ atomic operators during the sudden modulation of $a$ is negligible and the simple picture of the dynamical Casimir effect introduced in~\cite{noi_bastard} can be straightforwardly applied: the $\ah_\kk$, $\ahd_{-\kk}$ operators at $t=0^\pm$ right before and right after the jump are expanded in terms of the $\Lambdah_\kk(t=0)$, $\Lambdahd_{-\kk}(t=0)$ atomic operators using \eq{u_v} with the suitable $u_\kk$, $v_\kk$ Bogoliubov coefficients.
An explicit relation between the $\ah_\kk$, $\ahd_{-\kk}$ operators at $t=0^\pm$ is straightforwardly obtained by eliminating the $\Lambdah_\kk(t=0)$, $\Lambdahd_{-\kk}(t=0)$ operators. 
The evolution of the $\ah_\kk$, $\ahd_{-\kk}$ operators at later $t>0$ times reduces to the simple phase factor $\exp(\mp i\omega_\kk^+ t)$.

Once all these steps are combined together, one is finally led to the following compact relations~\cite{kiss}: 
\begin{eqnarray}
\ah_\kk(t)&=&\left[\frac{\eta^+_\kk+\eta^-_\kk}{2}\,\ah_\kk(0^-)+\frac{\eta^+_\kk-\eta^-_\kk}{2}\,\ahd_{-\kk}(0^-)\right]
\,e^{-i\omega_\kk^+t}  \nonumber \\ \eqname{jump1} \\ \nonumber\\
\ahd_{\kk}(t)&=&\left[\frac{\eta^+_\kk+\eta^-_\kk}{2}\,\ahd_{\kk}(0^-)+\frac{\eta^+_\kk-\eta^-_\kk}{2}\,\ah_{-\kk}(0^-)\right]
\,e^{i\omega_\kk^+t} \nonumber \\ \eqname{jump2}
\end{eqnarray}
where the $\eta^\pm_\kk$ coefficients are defined as $\eta^\pm_\kk=(\omega_\kk^+ / \omega_\kk^-)^{\pm 1/2}$ and the  Bogoliubov frequencies $\omega_\kk^\mp$ before and after the jump are evaluated using \eq{omega} with $a=a_{1,2}$. 

As $\eta^\pm_\kk \rightarrow 1$ for $E_\kk/\mu_{1,2}\gg 1$, the transformations (\ref{eq:jump1}-\ref{eq:jump2}) do not mix the $\ah_\kk$, $\ahd_{-\kk}$ operators for large values of $\kk$, which provides an {\em a posteriori} justification for the sudden jump condition $\sigma_t \mu_{1,2} \gg 1$. For the hydrodynamic modes with $E_\kk\ll \mu_{1,2}$, the mixing coefficient $\eta^\pm_\kk$ tends instead to a finite limiting value $(c_2/c_1)^{\pm 1/2}$.

The Bogoliubov mode occupation $n_\kk$ after the sudden change of scattering length is given by the formula
\begin{equation}
n_\kk=\frac{(\omega_\kk^+)^2 \, +\, (\omega_\kk^-)^2 }{2\,\omega_\kk^+\,\omega_\kk^-}\,n_\kk^{{\rm th},0}+
\frac{(\omega_\kk^+\,-\,\omega_\kk^-)^2}{4\,\omega_\kk^+\,\omega_\kk^-},
\eqname{n_k}
\end{equation}
while the anomalous average has the form
\begin{equation}
\mathcal{A}_\kk(t)=
\frac{1}{4}
\left(
\frac{\omega_\kk^+}{\omega_\kk^-}-\frac{\omega_\kk^-}{\omega_\kk^+}
\right)
\left(2n_\kk^{{\rm th},0}+1\right)
\,e^{-2i\omega_\kk^+ t}
\eqname{A_k}
\end{equation}
Note that the Bose distribution $n_\kk^{{\rm th},0}$ is here to be evaluated according to \eq{a+a} using the initial value $\omega_\kk^-$ of the Bogoliubov mode frequency. In agreement with \eq{aa}, the initial value of the anomalous average has been taken to be zero.
The terms in \eq{n_k} and \eq{A_k} proportional to the initial population $n_\kk^{{\rm th},0}$ account for the amplification of initial thermal excitations by the sudden jump in $a$, while the other terms describe the contribution due to the dynamical Casimir emission of Bogoliubov phonons out of the initial vacuum state.

It is interesting to evaluate \eq{n_k} and \eq{A_k} in the limit of a small change of $a(t)$, i.e. $a_2=a_1+\Delta a$ with $\Delta a \ll a_1$. In this limit, one has:
\begin{eqnarray}
n_\kk&\simeq&n_\kk^{{\rm th},0}+ \frac{\Delta \mu^2}{4(E_\kk+2\mu)^2}\,\left(2n_\kk^{{\rm th},0}+1\right)\eqname{n_pert} \\
\mathcal{A}_\kk(t)&\simeq&\frac{\Delta \mu}{2(E_k+2\mu)}\,\left(2n_\kk^{{\rm th},0}+1\right)\,e^{-2i\omega^+_\kk t}\,. \eqname{A_pert}
\end{eqnarray}
Here, the chemical potential variation is defined as $\Delta \mu=4\pi\hbar^2 (a_2-a_1)n_0 /m$. While the effect on the anomalous average is linear in $\Delta \mu$, the population change is quadratic and therefore much weaker. 
The difference is even more dramatic at $T>0$: while the population change is a (small) correction proportional to $\Delta \mu^2$ on top of the (large) initial thermal distribution, the anomalous average fully originates from the dynamical Casimir emission and is amplified by the initial thermal population.

\subsection{Physical discussion}

\begin{figure}[htbp]
\begin{center}
\includegraphics[width=0.95\columnwidth,angle=0,clip]{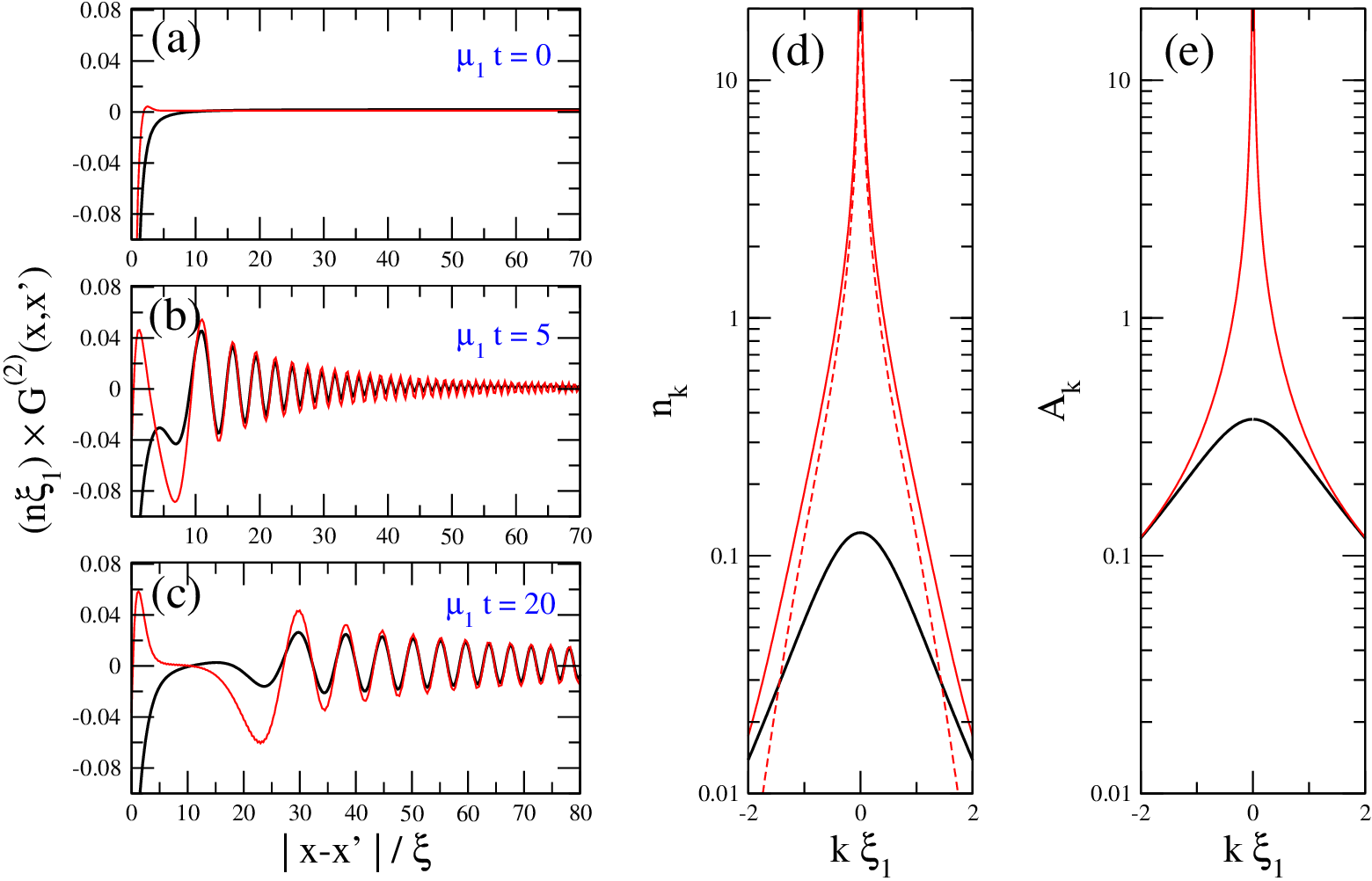}
\caption{
Panels (a-c): Time evolution of the density correlation function after a sudden jump of the scattering length from $a_1$ to $a_2=a_1/4$.  Different (a-c) panels refer to different evolution times after the jump, $\mu_1 t=0,\,5,\,20$. 
Panel (d): Bogoliubov mode occupation $n_\kk$ after the jump. 
Panel (e): anomalous average $|\mathcal{A}_\kk|$ after the jump. 
In all panels, thick black lines correspond to an initial $T=0$; thin red lines correspond to a finite initial temperature $T/\mu_1=0.5$. The dashed red line in panel (d) indicates the prediction of the adiabatic model, i.e. the initial population of the Bogoliubov mode $n_\kk^{{\rm th},0}$.
}
\label{fig:jump}
\end{center}
\end{figure}

This physics is illustrated in the plots of the Bogoliubov occupation and the modulus of the anomalous average shown in Fig.\ref{fig:jump}(d,e) for the $a_2/a_1=0.25$ case. For these plot, the exact formulas \eq{n_k} and \eq{A_k} have been used.

For a vanishing initial temperature $T=0$ (black line), both $n_\kk$ and $\mathcal{A}_\kk$ show a smooth peak centered at $\kk=0$ and a power-law tail that extends far on the high energy modes. As predicted by the analytical approximate formulas \eq{n_pert} and \eq{A_pert}, the anomalous average $\mathcal{A_\kk}$ is in modulus much larger than the occupation $n_\kk$. Note that the smooth peak would be replaced by a $1/k$ divergence if interactions in either the initial or final states were vanishing.

For a finite initial temperature $T>0$, the effect of the sudden variation of $a$ on $n_\kk$ is a very weak correction on top of the initial thermal distribution $n_\kk^{{\rm th},0}$. On the other hand, the $T=0$ contribution to the anomalous average due to the dynamical Casimir effect is strongly amplified by the initial thermal populaton.

The density correlation function is immediately obtained inserting (\ref{eq:n_k}-\ref{eq:A_k}) into the general expression \eq{G2anal}. 
A few snapshots after different evolution times are shown in Fig.\ref{fig:jump}(a-c) for the simplest case of a one-dimensional system. Exception made for some geometrical factors, the physics is however identical when two- or three-dimensional systems are considered.

For $t=0^+$, straightforward algebraic manipulations confirm that the sudden change of $a$ does not have an immediate effect on correlation function, $g^{(2)}(\xx-\xx';t=0^+)=g^{(2)}(\xx-\xx';t=0^-)$: The jump is in fact too rapid for the microscopic state of the atoms to respond. However, as this state is no longer an eigenstate of the system Hamiltonian with $a(t>0)=a_2$, a non-trivial evolution is observed on $g^{(2)}$ at later times $t>0$. 

Before the sudden change of $a$, the density correlation function is characterized by a dip around $x=x'$ due to the effect of atom-atom repulsive interactions. At finite temperature, this dip is less pronounced than at $T=0$, and starts being accompanied by a Hanbury-Brown and Twiss bump due to the thermal fluctuations~\cite{HBT}.

After the sudden change, the static central structure around $x=x'$ is somehow amplified by the change in population $n_\kk$ and, more importantly, by the increase in the Bogoliubov $u_\kk+v_\kk$ coefficient as a consequence of the reduced value of $a$.
At the same time, a system of moving fringes originates from $x=x'$ and propagates in the outwards direction as a consequence of the time-dependent anomalous average $\mathcal{A}_\kk(t)$.
At each time $t$, the fringe pattern is concentrated in the $|x-x'|\gtrsim c_2\, t$ region and shows a significant chirping in space, the external part of the pattern having a shorter wavelength than the inner part. As time goes on, the fringe pattern gets progressively stretched in space.

This peculiar fringe pattern has a very simple physical interpretation~\cite{calabrese}.
When the sudden change of $a$ occurs, pairs of entangled phonons are created at all spatial positions with opposite momenta $\pm k$.
As time goes on, these pairs propagate in opposite direction at a $k$-dependent group velocity $v_{2,k}$. 
As the group velocity $v_{2,k}$ of Bogoliubov modes is a growing function of $k$ which starts from $v_{2,k=0}=c_2$, the correlated pairs will be separated at a time $t$ by a distance equal or larger than $2 c_2 t$~\footnote{This behaviour is a consequence of the superluminal nature of the Bogoliubov dispersion \eq{omega}. This is to be contrasted with the linear or subluminal dispersions that were considered in several previous works.} 
More precisely, the modes that most contribute to the fringe pattern for a given $|x-x'|$ are the ones with a wavevector $\pm\bar{k}$ such that $v_{2,\bar{k}} \simeq |x-x'|/2t$, which are responsible for fringes of wavelength $1/\bar{k}$. 
In this semiclassical picture, the observed chirping is then a simple conseguence of the fact that higher $k$ modes propagate at a faster group velocity.

While the strongly chirped external region of the fringe pattern remains almost unaffected by a finite initial temperature, the long wavelength fringes at low $|x-x'|$ are substantially reinforced. This confirms our intuitive understanding of the fringe pattern: according to \eq{A_k} thermal enhancement is in fact concentrated into the low-$\kk$ modes which are responsible for the long wavelength fringes.

\section{Slow ramp}
\label{sec:slow}

The previous Section was devoted to the case of a sudden change of $a$ for which analytical expressions were available for the Bogoliubov mode amplitudes in the final state.
The more general case of an arbitrary dependence $a(t)$ requires a solution of the pair of ordinary differential equations (\ref{eq:ak}-\ref{eq:ak2}). 
In the present section we discuss the result of a numerical solution of these equations for the case of a smooth temporal dependence of the Erf form:
\begin{equation}
a(t)=\frac{a_1+a_2}{2}+\frac{a_2-a_1}{2}\;\textrm{Erf}\left( \frac{t-t_0}{\sigma_t}\right).
\end{equation}
where the change of scattering length from $a_1$ to $a_2$ takes place on a time scale $\sigma_t$.

\begin{figure}[htbp]
\begin{center}
\includegraphics[width=0.8\columnwidth,angle=0,clip]{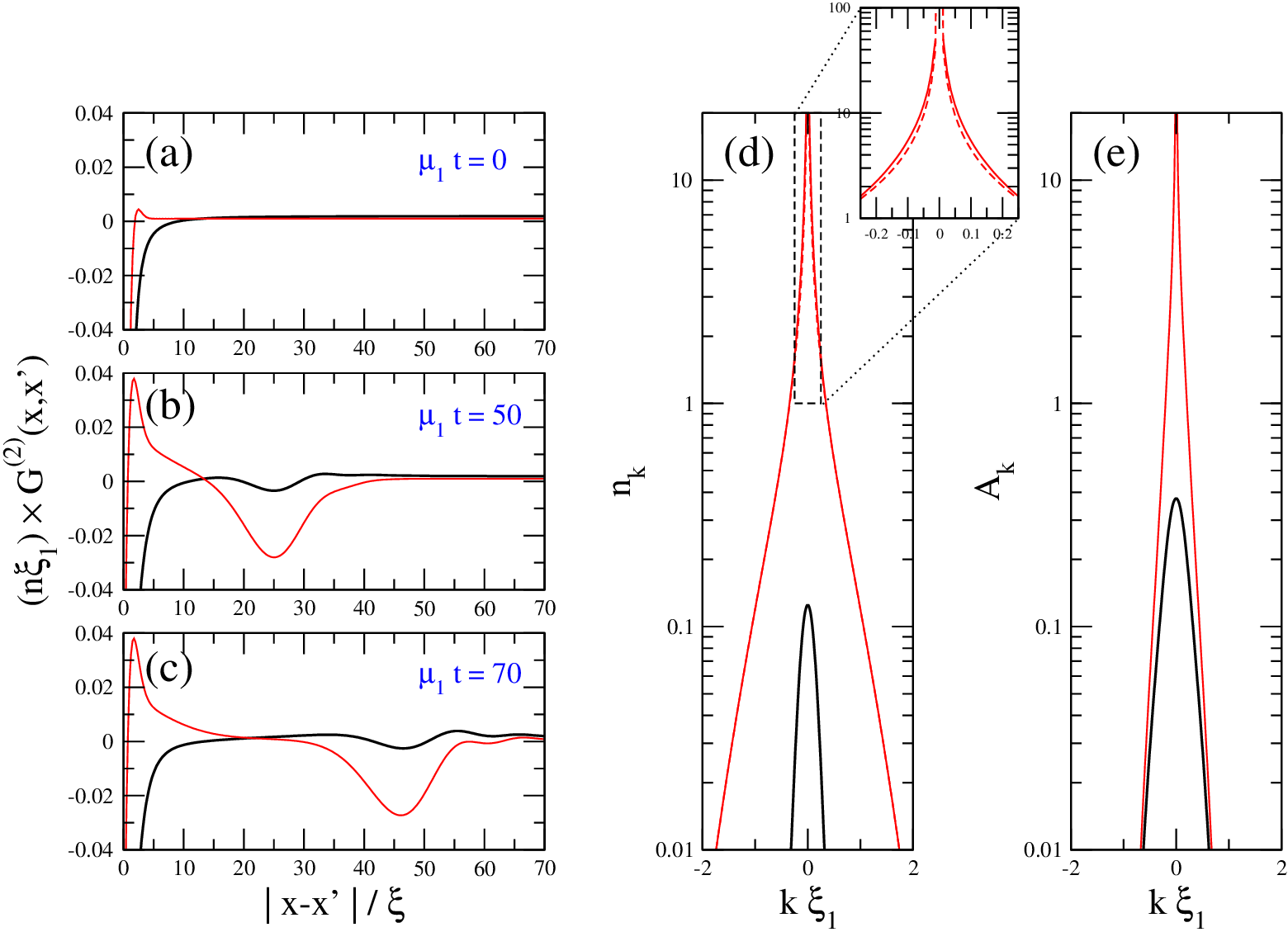}
\caption{
Panels (a-c): Time evolution of the density correlation function after a slow ramp of the scattering length from $a_1$ to $a_2=a_1/4$. The ramp follows an Erf shape with $\sigma_t=5/\mu_1$ centered at $t=5\sigma_t$. Different (a-c) panels refer to different evolution times after the jump, $\mu_1 t=0,\,50,\,70$. 
Panel (d): Bogoliubov mode occupation $n_\kk$ after the jump. A magnified view of the peak is given in the inset.
Panel (e): anomalous average $|\mathcal{A}_\kk|$ after the jump. 
In all panels, thick black lines correspond to an initial $T=0$; thin red lines correspond to a finite initial temperature $T/\mu_1=0.5$.  The dashed red line in (d) is the initial population $n_\kk^{{\rm th},0}$.}
\label{fig:smooth}
\end{center}
\end{figure}

As one can see in Fig.\ref{fig:smooth}(d,e), the main effect of a finite $\sigma_t$ is to introduce a ultraviolet cut-off to the Bogoliubov modes that are effectively excited during the modulation of $a$. 
All Bogoliubov modes with $\omega_\kk \sigma_t \gg 1$ experience in fact the modulation as adiabatic: as a consequence, the corresponding anomalous average $\mathcal{A}_\kk$ remains fully negligible and the population $n_\kk$ remains very close to its value before the ramp [dashed line in Fig.\ref{fig:smooth}(d)].

The density correlation function at different times after the slow modulation is shown in Fig.\ref{fig:smooth}(a-c) for the simplest case of a slow modulation rate as compared to the chemical potential, $\mu_{1,2} \sigma_t\gg 1$.
In this case, only the low-$k$, hydrodynamic modes result appreciably excited and the chirped fringes in the large $|x-x'|\gg 2 c_2 t$ region disappear from the fringe patterns shown in Fig.\ref{fig:smooth}(a-c). 
These are then characterized by a single negative peak that rigidly propagates at a speed $2c_2$ with almost no dispersion. 
A different point of view on this same phenomenology was recently presented in~\cite{calabrese}.
As one can see by comparing the thick black lines to the thin red ones in Fig.\ref{fig:smooth}(a-c), in this case the effect of a finite initial temperature reduces to an amplification of the propagating peak.

\section{Hydrodynamic limit}
\label{sec:hydro}

In the limit of a slow $\mu \sigma_t/\hbar \gg 1$ and weak $|\Delta \mu| \ll \mu$ jump, an analytical approximation can be obtained for the height and shape of the moving peak.
The idea is to use the sudden jump result \eq{A_pert} and then take into account the slow variation of $a(t)$ by means of a cut-off in the momentum integration of \eq{G2anal}: while the low-frequency modes $\omega_\kk\ll 1/\sigma_t$ experience the modulation as sudden, the high-frequency ones $\omega_\kk\gg 1/\sigma_t$ experience it as adiabatic.
The assumed condition $\mu \sigma_t/\hbar \gg 1$ implies that the momentum cut-off is at a wavevector $k_{\rm max}^t\simeq 1/c\sigma_t\ll 1/\xi$ well within the hydrodynamical regime for which $\omega_\kk\simeq c\,|\kk|$.

Including this cut-off as an additional exponential factor $\exp(-k/k_{\rm max}^t)$ in the integral of the zero-point contribution to \eq{G2anal}, we immediately get to the following expression for the moving peaks:
\begin{multline}
\delta g^{(2)}_{T=0}(x,x')\approx \frac{\hbar\,\Delta\mu}{4\pi m n\mu c}\times
\\ \times \left\{ \frac{\ell_t^2-(x-x'-2ct)^2}{[\ell_t^2+(x-x'-2ct)^2]^2}+\frac{\ell_t^2-(x-x'+2ct)^2}{[\ell_t^2+(x-x'+2ct)^2]^2}\right\}.
\eqname{fullpeakT=0}
\end{multline}
The peak value is at
\begin{equation}
\left.\delta g^{(2)}_{T=0}\right|_{\rm peak}\approx \frac{1}{4\pi\,(n\xi)\,(\mu\sigma_t/\hbar)^2}\,\frac{\Delta\mu}{\mu}
\eqname{deltaHD}
\end{equation}
and the width is determined by the cut-off as $\ell_t=1/k_{\rm max}^t=c\sigma_t$:
Physically, this value of the peak width can be understood as a consequence of the uncertainty $\sigma_t$ in the emission time, which reflects into a broadening $\ell_t=c\sigma_t$ of the correlation signal.

At a finite temperature $T>0$, one has to include the further contribution due to the amplified thermal fluctuations.
Depending on whether the temperature $k_B T$ is higher or lower than the effective cut-off energy $E_{\rm max}^t=\hbar c k_{\rm max}^t=\hbar /\sigma_t$ imposed by the slow ramp, the cut-off on the thermal contribution to \eq{G2anal} has to be imposed at $k_{\rm max}^{\rm th}=\min(k_{\max}^t,k_B T /\hbar c)$.
Including again this cut-off as an exponential factor $\exp(-k/k_{\rm max}^{\rm th})$, one gets the following expression for the moving peaks:
\begin{multline}
\delta g^{(2)}_{{\rm th}}(x,x')\approx \frac{\Delta\mu\,k_BT\,\ell_{\rm th}}{2\mu^2 n}\times
\\ \times \left[ \frac{1}{(x-x'-2ct)^2+\ell_{\rm th}^2}+ \frac{1}{(x-x'+ 2ct)^2+\ell_{\rm th}^2}\right].
\eqname{fullpeakT}
\end{multline}
The width of the thermal peaks is set by the cut-off $\ell_{\rm th}=1/k_{\rm max}^{\rm th}$. At low temperature $k_B T < E_{\rm max}^t$, the width is enlarged to $\ell_{\rm th}= \hbar c/k_B T$ as a consequence of the finite correlation length of thermal density fluctuations in the initial state. At high temperature $k_B T > E_{\rm max}^t$, the width is again dominated by the ramp time effect as in the $T=0$ case.

Depending on whether $k_B T \gtrless E_{\rm max}^t$, the height of the thermal peaks is either:
\begin{equation}
\left.\delta g^{(2)}_{{\rm th}}\right|_{{\rm peak,\,high-} T}\approx \frac{1}{4\,(n\xi)\,(\mu\sigma_t/\hbar)}\,\frac{\Delta\mu\;k_BT}{\mu^2}
\eqname{peakThigh}
\end{equation}
or
\begin{equation}
\left.\delta g^{(2)}_{{\rm th}}\right|_{{\rm peak,\,low-}T}\approx \frac{1}{4\,(n\xi)}\,\frac{\Delta\mu\,(k_BT)^2}{\mu^3}. \eqname{peakTlow}
\end{equation}

Even though the parameters used in Fig.\ref{fig:smooth} are on the edge of the validity domain of the hydrodynamic approximation, the analytical formulas discussed in the present section turns out to be in reasonable quantitative agreement with the numerical results.

\section{Comparison with black hole calculations}
\label{sec:compMC}

\begin{figure}[htbp]
\begin{center}
\includegraphics[width=0.95\columnwidth,angle=0,clip]{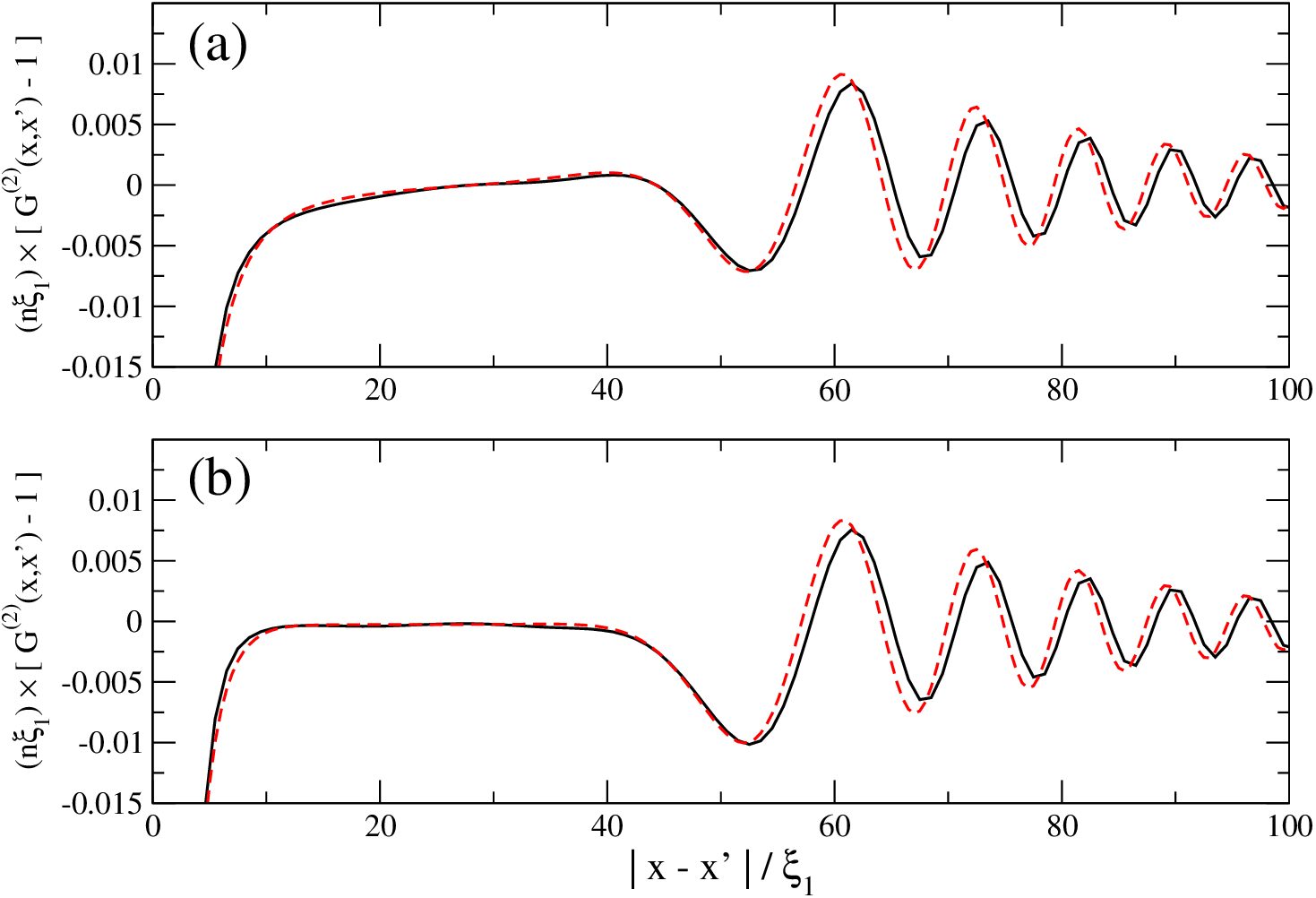}
\caption{Comparison of the Bogoliubov prediction \eq{G2anal} for the density correlation function (thick black line) with the result of the numerical simulations of~\cite{noi_NJP} (thin red dashed line). The numerical lines are cuts of $G^{(2)}(x,x')$ along a $x+x'=2x_0$ line with $x_0/\xi_{1,2}\gg 1$ well inside the acoustic black hole region; all the field of view is within the acoustic black hole region.
The scattering length is brought from $a_1$ to $a_2=a_1/4$ with a Arctan temporal dependence on a time scale $\sigma_t \mu_1 =0.5$. The correlation functions are evaluated a time $\mu_1 t=50$ after the change of $a$.
The upper (a) panel is for $T=0$. The lower (b) panel is for a finite $k_B T/\mu_1=0.1$. 
}
\label{fig:compMC}
\end{center}
\end{figure}

In the previous sections we have investigated in detail the density correlations that appear as a consequence of a modulation of the atomic scattering length $a(t)$.
One of the motivations of the present work was to fully understand some unexpected transient features that were observed in the numerical experiment of~\cite{noi_NJP}, namely a system of moving fringes that appear inside the acoustic black hole as soon as the horizon is created and then rapidly leave the field of view.
As the acoustic horizon was created by suddenly ramping down the atomic scattering length in a full half space and a quantitatively identical system of fringes was observed in a spatially homogeneous system, an interpretation was put forward in terms of dynamic Casimir effect.
In this Section, we confirm this interpretation by performing a quantitative comparison of the numerical results of~\cite{noi_NJP} to the predictions of the Bogoliubov model that we have discussed in the previous Sections. To this purpose, the same Arctan-shaped ramp of $a(t)$ that was used in the numerical calculations has to be implemented in the Bogoliubov calculation: the results of the comparison are shown in Fig.\ref{fig:compMC}. The agreement between the two calculations is remarkable, which firmly confirms our initial interpretation.

\section{Quasi-periodic modulation}
\label{sec:periodic}

\begin{figure}[htbp]
\begin{center}
\includegraphics[width=0.9\columnwidth,angle=0,clip]{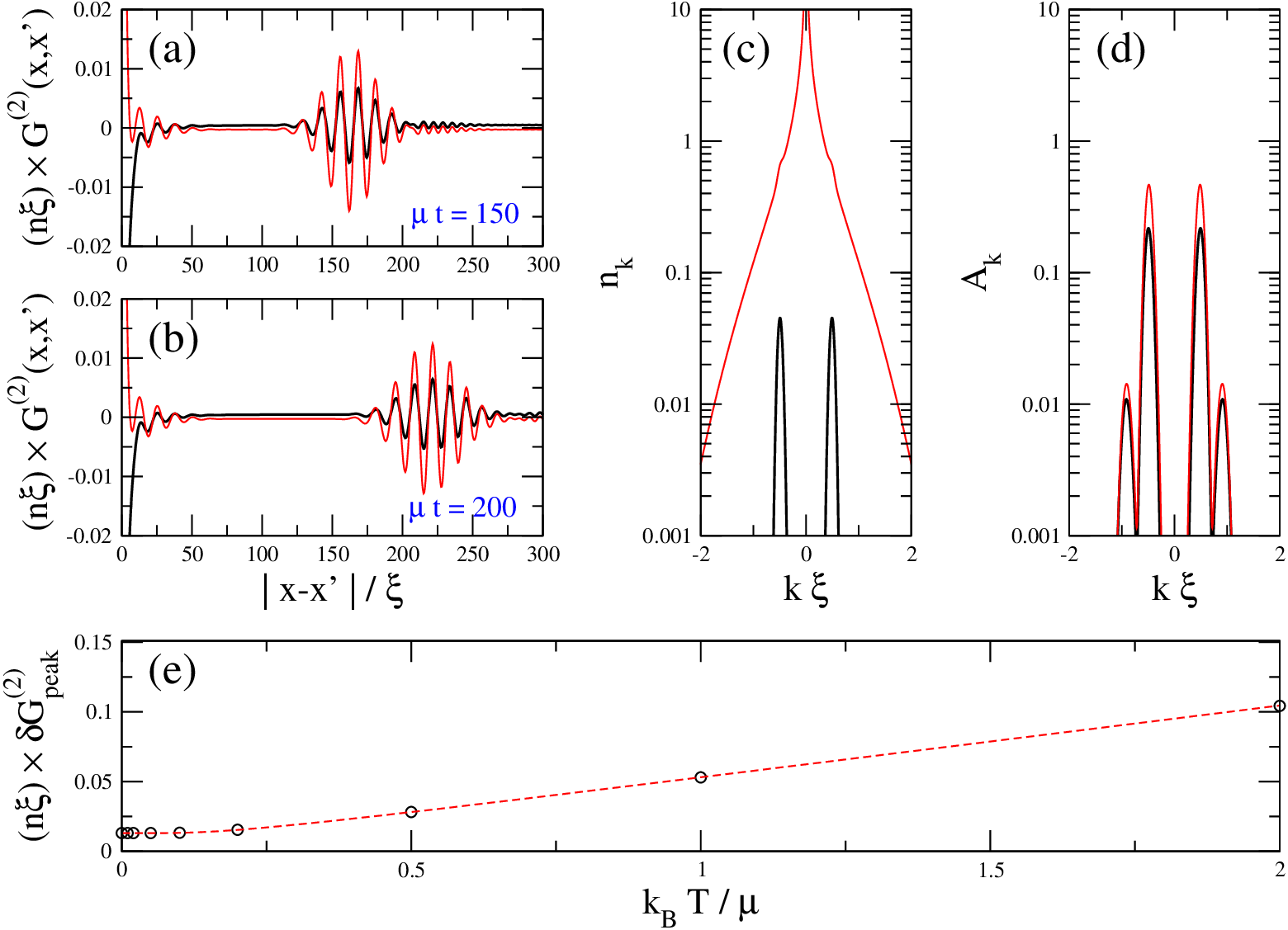}
\caption{Panels (a,b): Time evolution of the density correlation function after an oscillating modulation of the scattering length of amplitude $\delta a/a=0.1$, carrier frequency $\hbar\omega_0/\mu=1$, and Gaussian envelope of duration $\omega_0 T=10$. 
The two panels refer to different evolution times after the peak of the modulation, $\mu t=150$ (a) and $200$ (b).
Panel (c): Bogoliubov mode occupation $n_\kk$ after the modulation. Panel (d): anomalous average $|\mathcal{A}_\kk|$ after the modulation.  In all (a-d) panels, black lines are for a zero initial temperature $T=0$; redlines are for an initial temperature $k_B T/\mu=0.5$. 
Panel (e): Peak value of the fringe amplitude as a function of the initial temperature for a given quasi-periodic excitation sequence. Black circles: numerical integration of  (\ref{eq:ak}-\ref{eq:ak2}). Dashed red line: fit of the points with a thermal law $(1+2n_\kk^{{\rm th},0})$.
}
\label{fig:periodic}
\end{center}
\end{figure}

A narrow window of $\kk$ modes can be specifically addressed by using a periodic modulation of the scattering length in time: according to the form \eq{Ham_pert} of the system Hamiltonian, a weak perturbation at frequency $\omega_0$ is in fact able to effectively excite those pairs of Bogoliubov modes that satisfy the resonance condition
\begin{equation}
\omega_0=\omega_\kk+\omega_{-\kk}.
\end{equation}
This physics is illustrated in Fig.\ref{fig:periodic}, where we show the result of a numerical integration of (\ref{eq:ak}-\ref{eq:ak2}) under a sinusoidal modulation of $a(t)$ with a Gaussian temporal envelope: the mixing of the $\ah_\kk$ and $\ahd_{-\kk}$ operators is limited to a small range of $\kk$ vectors and results in very peaked shapes of the Bogoliubov mode occupation $n_\kk$ and of the anomalous average $\mathcal{A}_\kk$ [panels (c,d)].

The pair of weaker peaks that appears in the anomalous average $\mathcal{A}_\kk$ at larger values of $k$ is due to second-order processes in the modulation amplitude. This interpretation is confirmed by the scaling of the peak amplitude as $\delta a^2$ and by the position of the peak that satisfies the second-order resonance condition $\omega_\kk+\omega_{-\kk}=2\omega_0$.
For larger modulation amplitudes, higher order peaks would appear at $k$ values satisfying higher-order resonance conditions of the form $\omega_\kk+\omega_{-\kk}=N\,\omega_0$, $N$ being a generic positive integer number.

Correspondingly to the dominant pair of peaks, the density correlation function [panels (a,b)] shows a periodic fringe pattern of wavelength $2\pi/k$ that travels away from $|x-x'|=0$ at the group velocity $2v_{2,k}$. The envelope of the fringe pattern follows the envelope of the oscillating $a(t)$ modulation.

At a finite temperature (thin red lines), the resonance peaks in the Bogoliubov mode occupation $n_\kk$ are almost completely hidden by the thermal component, but remain perfectly visible and even reinforced in the anomalous average $\mathcal{A}_\kk$. 
Correspondingly, the moving fringe pattern experiences an overall amplification without any significant distortion of the oscillating shape nor of its envelope.

The simple relations \eq{G2anal} and \eq{A_k_gen} that relate the density correlation at the end of the modulation sequence to the initial thermal occupation of the mode can be exploited as a simple way to precisely measure the temperature of the system in an almost non-destructive way. 
This proposal extends an original suggestion of~\cite{tozzo} to measure the temperature of a Bose-Einstein condensate using a parametric modulation of some parameter: as illustrated in Fig.\ref{fig:periodic}(c), looking at the density correlation rather than at the Bogoliubov mode occupation has the significant advantage that the interesting signal is not hidden by a broad thermal pedestal. In contrast to the case of a single jump discussed~\cite{Demler}, a periodic modulation is able to concentrate the interesting signal into a single Bogoliubov mode and, more remarkably, to make it significantly stronger without distorting it.

As long as the applied modulation is weak enough for nonlinear and saturation effect beyond Bogoliubov theory to be negligible, the observed signal is in fact proportional to $2n_\kk^{{\rm th},0}+1$. To clarify this statement, the peak value of the fringe amplitude is plotted in Fig.\ref{fig:periodic}(e) as a function of the initial temperature. The points are the result of a numerical integration of  (\ref{eq:ak}-\ref{eq:ak2}), the dashed line is a fit using the known thermal dependence: provided a suitably low-energy mode is used, the peak value of the fringe amplitude is proportional to the system temperature.

\section{Reinforcing the density correlation}
\label{sec:trick}


A critical issue in view of an experimental verification of the conclusions of the present paper as well as of the predictions of~\cite{noi_NJP} is the actual value of the density correlation signal that one is to detect: given its relatively small value, methods to reinforce it can be of crucial importance. 
In the present section we apply to the dynamical Casimir effect a diagnostic trick that was recently used to characterize phase fluctuations of a quasi-condensate in a strongly one-dimensional geometry e.g. in~\cite{dettmer} and that was recently put forward in the context of observing the analog Hawking radiation~\cite{Cornell}.
The efficiency of this tool to measure the microscopic properties of low-dimensional many-body systems was recently discussed in~\cite{Demler}. A related idea was presented in~\cite{gardiner} with the purpose of amplifying the signal of analog cosmological particle production.

\begin{figure}[htbp]
\begin{center}
\includegraphics[width=0.95\columnwidth,angle=0,clip]{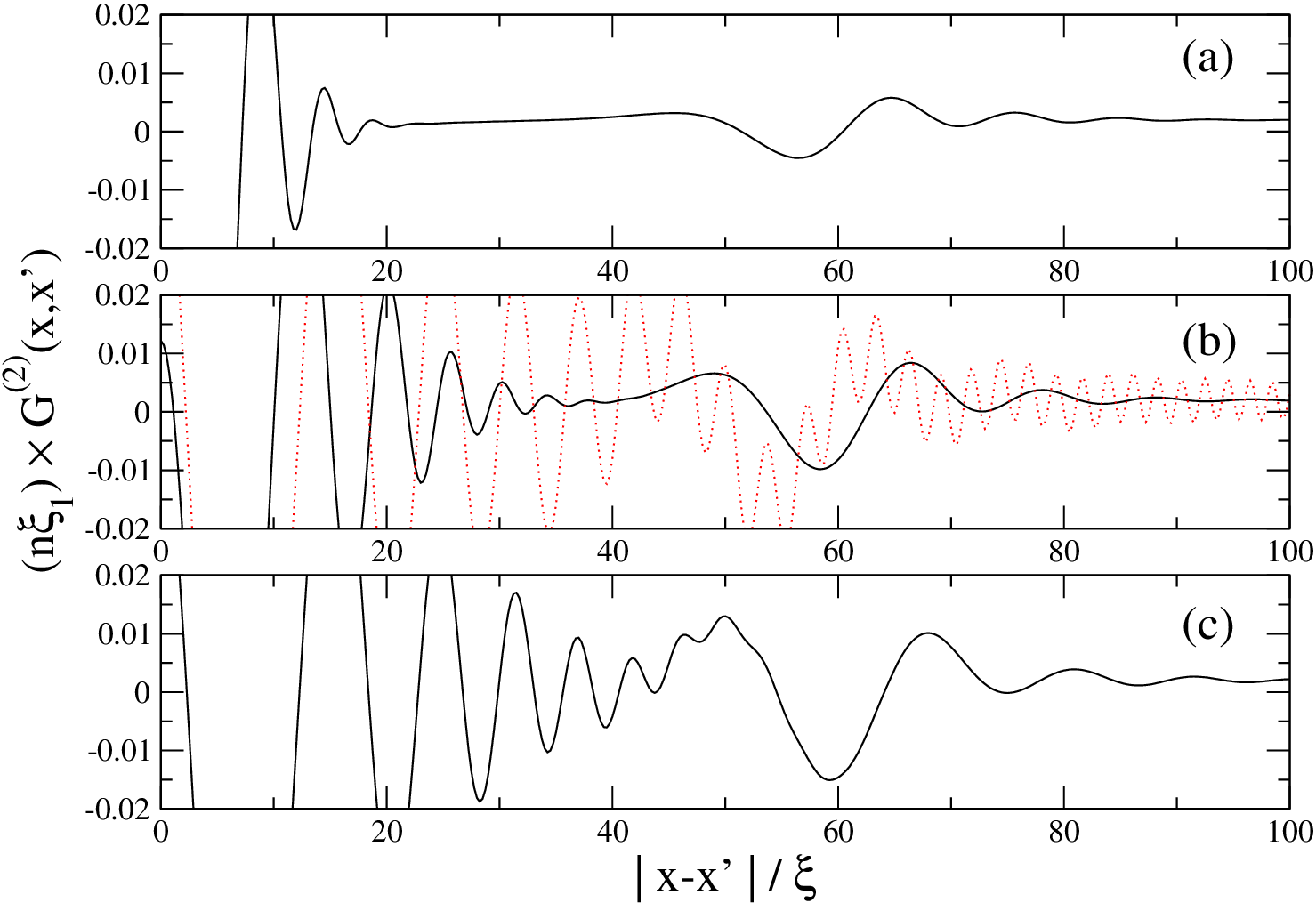}
\caption{Density correlation function after a slow ramp of the scattering length from $a_1$ to $a_2=a_1/4$ as in Fig.\ref{fig:smooth}, a faster switch-off of the scattering length to $a_f=0$ within $\sigma_t'\ll \sigma_t$, and a final time interval $t_{\rm free}$ of ballistic, non-interacting evolution.
Different panels (a-c) correspond to different free evolution times $\mu_1\,t_{\rm free}=0$ (a) $5$ (b), and $10$ (c). The switch-off of the scattering length to $a_f=0$ is performed at $\mu_1 t=70$ within a time $\mu_1\sigma'_t=1$ (black lines). For comparison, the case of a sudden switch-off ($\sigma_t'=0$) is shown as a red dotted line in (b). 
}
\label{fig:trick1}
\end{center}
\end{figure}

In agreement to the Goldstone theorem, long wavelength phonons have a mostly phase-like character and a very weak component of density fluctuation~\cite{castin}: a quick switch-off of the the interactions shortly before measuring the density correlations can then be used in order to reinforce the signal by converting phase fluctuations into density fluctuations.
The efficiency of this trick is illustrated in Fig.\ref{fig:trick1}: the density correlation signal is plotted for different values of the ballistic expansion time $t_{{\rm fin}}$ between the switch-off of $a$ and the actual measurement. During this time, the original signal gets amplified by a significative factor. 

Note that in order to avoid a substantial emission of high-$k$ particles and the consequent appearance of fast oscillations in the density correlation pattern, the switch-off time $\sigma'_t$ can not be chosen too short. This point is illustrated in Fig.\ref{fig:trick1}(b) where the signal obtained with a sudden final jump is shown for comparison as a red dotted line: the importance of a careful choice of $\sigma_t'$ is apparent. 

An analytical understanding of the physical origin of the different features that appear in the density correlation function after the second jump is the subject of the next subsection.

\subsection{Hydrodynamic model}

Analytical expressions can be obtained in the case in which the second jump brings the scattering length to a finite final value $a_f$ and both jumps are performed on a time-scale $\sigma_t,\sigma_t'$ long as compared to the chemical potential, $\sigma_t,\sigma_t' \gg 1/\mu_1$. Under these assumptions, the analytical technique introduced in Sec.\ref{sec:hydro} can be generalized to the case of a two-jump modulation sequence.

\begin{figure}[htbp]
\begin{center}
\includegraphics[width=0.95\columnwidth,angle=0,clip]{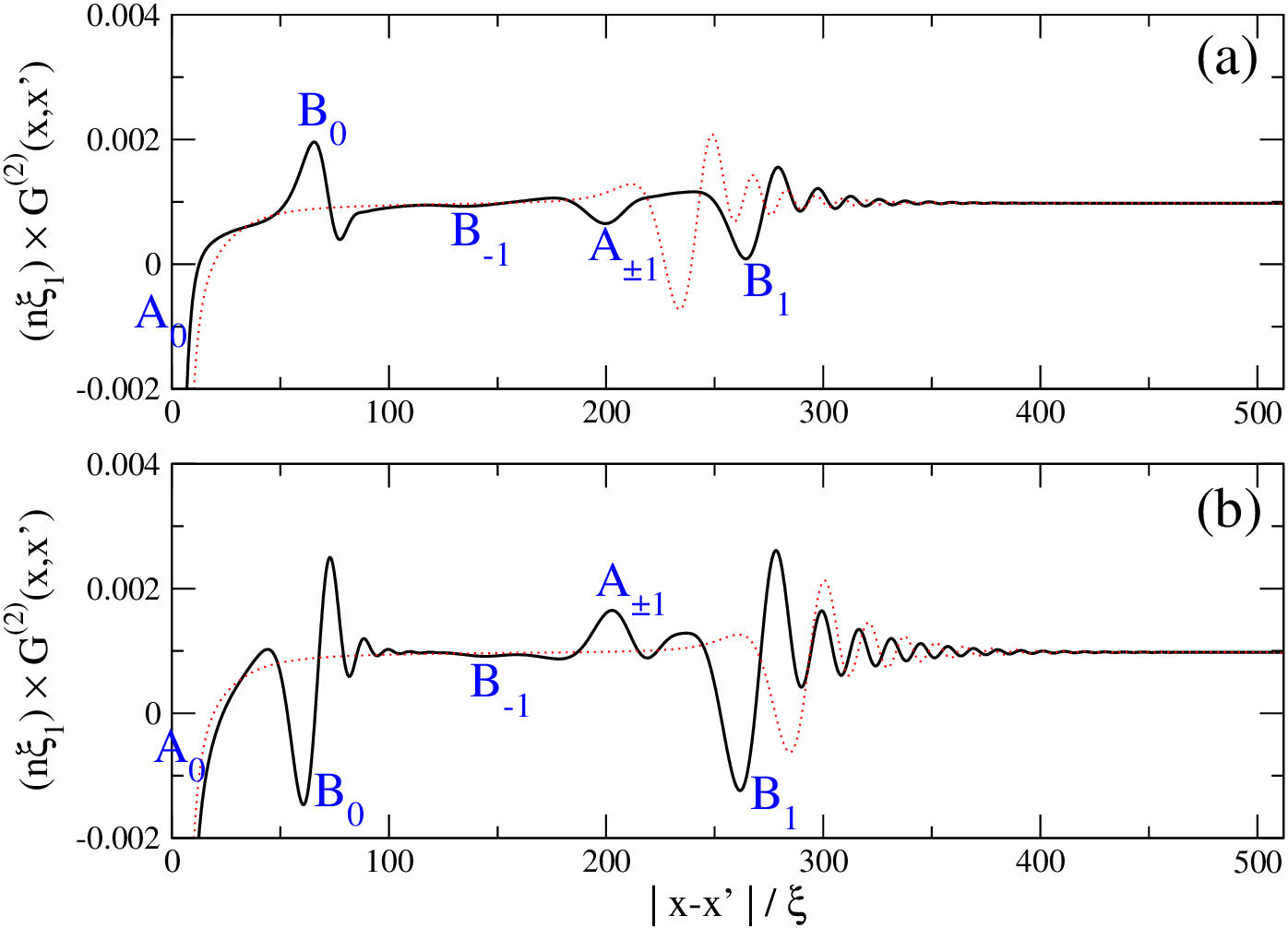}
\caption{Density correlation function after a two-jump sequence.
Jump times $\sigma_t,\sigma_t'=8/\mu_1$. Delay time $t_{del}=100/\mu_1$.
Upper panel (a): $a_2=a_1/4$, $a_f=a_1$, observation time $t=t_{\rm free}+t_{\rm del}=270/\mu_1$. 
Lower panel (b): $a_2=a_1/4$, $a_f=a_1/8$, observation time $t=t_{\rm free}+t_{\rm del}=320/\mu_1$.
The blue labels indicate the terms in the two-sudden-jumps analytical model that correspond to each feature.
Red lines indicate the density correlation function in the absence of second jump, $a_f=a_2$.
}
\label{fig:trick2}
\end{center}
\end{figure}

\begin{figure}[htbp]
\begin{center}
\includegraphics[width=0.95\columnwidth,angle=0,clip]{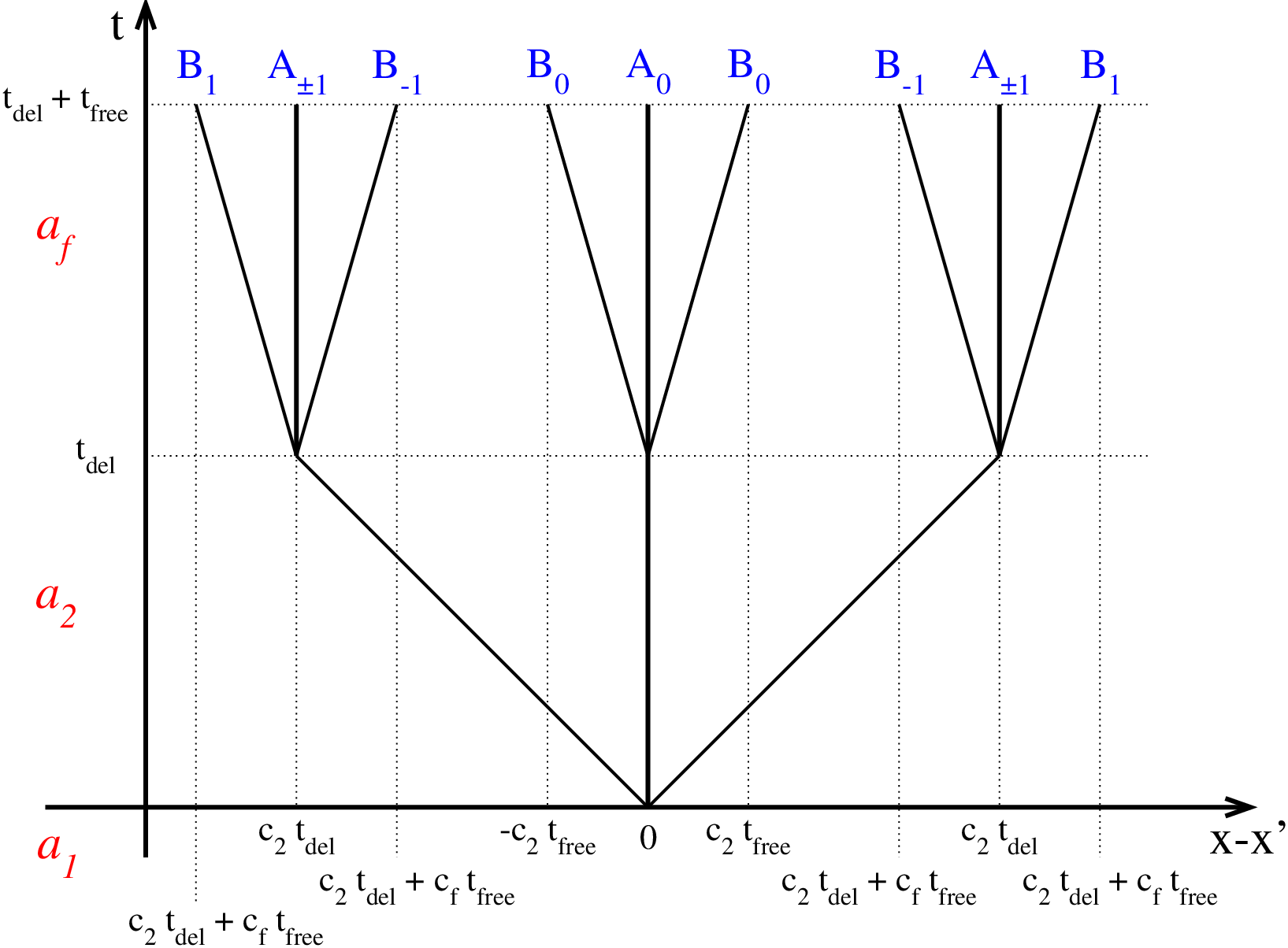}
\caption{Scheme of the spatial position of the different features predicted by \eq{G2hydro_2jumps} for a configuration $a_1>a_2>a_f$ inspired to  Fig.\ref{fig:trick2}(b).}
\label{fig:sketch}
\end{center}
\end{figure}

An explicit forms for the Bogoliubov coefficients a time $t_{\rm fin}$ after the end of the two-jump modulation sequence can be obtained by repeatedly applying two transformations of the form (\ref{eq:jump1}-\ref{eq:jump2})~:
\begin{eqnarray}
C_{\kk,+}&=& \left[ 
C_{\kk,+}^{(2)} C_{\kk,+}^{(1)}\,e^{-i\omega_\kk^{(2)}\,t_{\rm del}}+ \right. \nonumber \\
&+& \left. C_{\kk,-}^{(2)} C_{\kk,-}^{(1)}\,e^{i\omega_\kk^{(2)}t_{\rm del}}
\right]\,e^{-i\omega_\kk^{(f)}\,t_{\rm fin}}
\eqname{2C+} \\
C_{\kk,-}&=& \left[ 
C_{\kk,+}^{(2)} C_{\kk,-}^{(1)}\,e^{-i\omega_\kk^{(2)}\,t_{\rm del}}+ \right. \nonumber \\
&+&\left. C_{\kk,-}^{(2)} C_{\kk,+}^{(1)}\,e^{i\omega_\kk^{(2)}t_{\rm del}}
\right]\,e^{-i\omega_\kk^{(f)}\,t_{\rm fin}}
\eqname{2C-}
\end{eqnarray}
Here, $t_{\rm del}$ is the time interval between the jumps and $t_{\rm fin}$ is the time interval between the second jump and the actual measurement. $\omega_\kk^{(1,2,f)}$ are the Bogoliubov dispersions before the first jump, after the first jump, and after second jump, respectively. 
The single jump Bogoliubov coefficients $C_{\kk,\pm}^{(1,2)}$ are defined according to \eq{jump1} and \eq{jump2} as:
\begin{eqnarray}
C_{\kk,\pm}^{(1)}= \frac{1}{2}\left(\sqrt{\frac{\omega_{\kk}^{(2)}}{\omega_{\kk}^{(1)}} }\pm \sqrt{\frac{\omega_{\kk}^{(1)}}{\omega_{\kk}^{(2)}} }\right) \\
C_{\kk,\pm}^{(2)}= \frac{1}{2}\left(\sqrt{\frac{\omega_{\kk}^{(f)}}{\omega_{\kk}^{(2)}} }\pm \sqrt{\frac{\omega_{\kk}^{(2)}}{\omega_{\kk}^{(f)}} }\right)
\end{eqnarray}
The corresponding density correlation function is then obtained by inserting these formulas into the general formula \eq{G2anal} and imposing a suitable cut-off to the integrals at $k_{\rm max}^t=1/\ell_t$. Limiting ourselves to the simplest $T=0$ case, some straightforward algebra leads to the final result:
\begin{multline}
\delta g^{(2)}(X=x-x')=\frac{\hbar}{2\pi m n c_f}
\left\{
A_1\,F_{\ell_t}[X-2c_2 t_{\rm del}]+ \right. \\
+A_0\,F_{\ell_t}[X]
+A_{-1}\,F_{\ell_t}[X+2c_2 t_{\rm del}]+ \\
+B_1\,F_{\ell_t}[X-2c_f t_{\rm free}-2c_2t_{\rm del}]+ \\
+B_{0}\,F_{\ell_t}[X-2c_f t_{\rm free}] \\
+ B_{-1}\,F_{\ell_t}[X-2c_f t_{\rm free}+2c_2t_{\rm del}]+ \\
\left. +(X\leftrightarrow -X)
\right\}.
\eqname{G2hydro_2jumps}
\end{multline}
The amplitudes have the following expressions in terms of the Bogoliubov operators of the two jumps:
\begin{eqnarray}
A_{\pm 1}&=& C_{\kk,+}^{(2)}\,C_{\kk,-}^{(1)}\,C_{\kk,-}^{(2)}\,C_{\kk,+}^{(1)}
\eqname{A1} \\
A_0&=& |C_{\kk,+}^{(2)}\,C_{\kk,-}^{(1)}|^2+ |C_{\kk,-}^{(2)}\,C_{\kk,+}^{(1)}|^2 
\\
B_1&=&C_{\kk,+}^{(2)}\,C_{\kk,+}^{(1)}\,C_{\kk,+}^{(2)}\,C_{\kk,-}^{(1)} 
\\
B_{0}&=&C_{\kk,+}^{(2)}\,C_{\kk,+}^{(1)}\,C_{\kk,-}^{(2)}\,C_{\kk,+}^{(1)}+ \nonumber \\
&+&C_{\kk,-}^{(2)}\,C_{\kk,-}^{(1)}\,C_{\kk,+}^{(2)}\,C_{\kk,-}^{(1)}
\\
B_{-1}&=&C_{\kk,-}^{(2)}\,C_{\kk,-}^{(1)}\,C_{\kk,-}^{(2)}\,C_{\kk,+}^{(1)}
, \eqname{B-1}
\end{eqnarray}
and the function $F_\ell(x)$ is defined as
\begin{equation}
F_\ell(x)=\frac{\ell^2-x^2}{[\ell^2+x^2]^2}.
\end{equation}

As a consequence of the interference between the different terms of (\ref{eq:2C+}-\ref{eq:2C-}), a number of peak/dips appear in the final result \eq{G2hydro_2jumps} and have a peculiar evolution as a function of $t_{\rm free}$.
An illustration of this physics is shown in Fig.\ref{fig:trick2}: even though significantly distorted by effects beyond hydrodynamics, all the features are clearly recognizable. Labels refer to the corresponding amplitudes defined in eqs.(\ref{eq:A1}-\ref{eq:B-1}) and schematically illustrated in Fig.\ref{fig:sketch}.

The $A_0$ feature provides a slight modification of the many-body dip.
The standard dynamical Casimir effect by the second jump is responsible for the feature $B_0$ that emerges from the many-body dip at $x=x'$ at travels at a speed $2c_f$. In agreement with \eq{deltaHD}, its sign depends on the sign of the second jump $\Delta a_f=a_f-a_2$.

The dynamical Casimir feature that was visible at $2 c_2 t_{\rm del}$ before the second jump splits into three features $B_{-1}$, $A_{\pm 1}$ and $B_1$ that travel away at speeds respectively equal to $-2c_f$, $0$, $2c_f$. 
In the absence of second jump (i.e. for $a_f=a_2$, dashed red line in Fig.\ref{fig:trick2}), only the $B_1$ survives with a finite amplitude, though at a slightly shifted position as a consequence of the unchanged sound velocity. 
All other $B_{-1}$, $A_{\pm 1}$ features instead vanish as a consequence of the $C^{(2)}_{\kk,-}=0$ condition.
Before the jump, the height of the $B_1$ feature is proportional to $C_{\kk,+}^{(1)}\,C_{\kk,-}^{(1)}/c_2$. 
At the jump, it gets multiplied by a factor approximately equal to:
\begin{equation}
\eta=\left[\frac{1}{2}\left(1+\frac{c_2}{c_f}\right)\right]^2.
\end{equation}
As expected, this factor is larger than $1$ as soon as the second jump corresponds to a decrease in the scattering length $a_f<a_2$. In particular, it becomes very large when $a_f$ is brought to a very small value $a_f\ll a_2$.

Even though these analytical considerations are limited to the hydrodynamic regimes, they provide an useful qualitative guidelines to interpret the full numerical results shown in Fig.\ref{fig:trick1} and \ref{fig:trick2}.

\section{Conclusions}
\label{sec:conclu}

In this paper, we have presented a general theory of the density fluctuations that appear in an atomic Bose-Einstein condensate as a consequence of a temporal modulation of the atomic scattering length. Different regimes have been identified as a function of the time scale and the temporal shape of the modulation. A physical picture in terms of the dynamical Casimir emission of pairs of entangled phonons has provided an intuitive explanation of the results. Simple analytical formulas have been obtained in the most remarkable limiting cases. 
Excellent agreement is found with the quantum Monte Carlo calculation of Ref.\cite{noi_NJP}, which {\em a posteriori} confirms the physical interpretation of the numerical data. The efficiency of a recently proposed strategy~\cite{Cornell} to reinforce the experimental signal is discussed and quantitatively validated.  Possible applications to the thermometry of ultracold atomic gases are pointed out. 

We are grateful to C. Tozzo, F. Dalfovo, E. Cornell, and P. Calabrese for stimulating exchanges and discussions. A long-lasting collaboration with C. Ciuti and S. De Liberato on the Dynamical Casimir Effect is warmly acknowledged.

\end{document}